\documentclass[twocolumn]{biophys}
\usepackage{amssymb,amsfonts,amsthm,bm}
\usepackage[round,numbers,sort&compress]{natbib}
\usepackage[ansinew]{inputenc}
\usepackage{comment}
\usepackage{multirow}

\jno{kxl014}
\gridframe{N}
\cropmark{Y}
\doi{}

\begin{document}

\title{Characterizing the folding core of the cyclophilin A --- cyclosporin A complex II: improving folding core predictions by including mobility}

\author{J. W. Heal$^{\ast}$$^{\dagger}$, S. A. Wells$^{\ddagger}$, R. B. Freedman$^{\S}$ and R. A. R\"{o}mer$^{\P}$}
\address{$^{\dagger}$MOAC Doctoral Training Centre and Institute for Advanced Study, University of Warwick, Coventry, UK. $^{\ddagger}$Department of Physics, University of Bath, Bath, UK. $^{\S}$School of Life Sciences, University of Warwick, Coventry, UK. $^{\P}$Department of Physics and Centre for Scientific Computing, University of Warwick, Coventry, CV4 7AL, UK.}

\begin{abstract}
{Determining the folding core of a protein yields information about its folding process and dynamics. The experimental procedures for identifying the amino acids which make up the folding core include hydrogen-deuterium exchange and $\Phi$-value analysis and can be expensive and time consuming.
As such there is a desire to improve upon existing methods for determining protein folding cores theoretically. Here, we use a combined method of rigidity analysis alongside coarse-grained simulations of protein motion in order to improve folding core predictions for unbound CypA and for the CypA-CsA complex. 
We find that the most specific prediction of folding cores in CypA and CypA-CsA comes from the intersection of the results of static rigidity analysis, implemented in the {\sc First} software suite, and simulations of the propensity for flexible motion, using the {\sc Froda} tool.
%
}
{$Revision:$, compiled \today}
{*Correspondence: jack.heal@bristol.ac.uk \\
Address reprint requests to Jack Heal, University of Bristol, School of Chemistry, Cantock's Close, Bristol, BS8 1TS, UK.} 
\end{abstract}

\maketitle

\markboth{Heal et al.}{The CypA-CsA folding core II}

\section*{INTRODUCTION}

\section*{Predicting folding cores}

In Ref.\ \cite{HeaRBF14} we determined the experimental folding core for the complex between the multifunctional protein cyclophilin A (CypA) and the immunosuppressant drug cyclosporin A (CsA) using hydrogen-deuterium exchange NMR (HDX). We showed that the established theoretical method of rigidity analysis implemented using the {\sc First} software provides a reasonable prediction of the experimental folding core  \citep{HesRTK02,RadB04,RadAIK04}. In addition to this, we generated the {\sc First} folding core (FIR) for unbound CypA, which broadly agreed with the previously published data from HDX experiments. Whilst impressive, the {\sc First} method is not perfect and does not capture the changes in the HDX folding core observed upon ligand binding.
Indeed, FIR is larger in the absence of the ligand, a result which contrasts with what is observed experimentally. That there exists some discrepancy between theory and experiment could be expected since {\sc First} is used to make inferences about protein flexibility based upon a {\it static} crystal input structure only. In addition to flexibility, HDX is dependent upon the surface exposure of the exchanging protons as well as protein motion \citep{Woo93,LiW99}.
With this in mind, we now seek to improve upon the predictive power of the rigidity analysis method by incorporating information on the surface exposure and dynamics of the protein.
We change the hydrogen bond (HB) and hydrophobic tether (HP) networks so that these interactions are absent for surface atoms. With this modification we enhance the flexibility of the protein surface so that the remaining rigid residues may better correlate with the experimental folding core.
Building upon this, we model the prospective dynamics of the protein using coarse-grained simulations. With these techniques we are able to probe the propensity for large-amplitude motions which may only be accessed over long timescales, such as the timescales of our HDX experiments.

\section*{Coarse-grained simulations of protein motion}

The study of protein motion is of fundamental importance in structural biology due to its intimate link with protein function \citep{HenK07}.
The most prevalent method for modeling protein motion computationally is through molecular dynamics (MD) simulations \citep{GohT06}, yielding detailed atomic trajectories \citep{KarP90}.
MD simulations are known to be computationally intensive; yet some of the most biologically relevant motions take place on the ms--s timescale which remains inaccessible to MD at present \citep{WelMHT05}.
Methods of coarse graining may be applied to reduce computational demand and to enable simulation of  larger amplitude protein motion \citep{Cle08}.
The constrained geometric simulation software {\sc Froda} (framework rigidity optimized dynamic algorithm) \label{froda} is a module within {\sc First} which rapidly simulates possible protein motion \citep{WelMHT05}.  It has been used to investigate protein-protein docking problems \citep{JolWHT06} as well as substrate recognition and conformational changes \citep{MacNBC07}. Large-amplitude motions in proteins with hundreds or thousands of residues can be rapidly explored using a desktop computer on a timescale of minutes \citep{JimFRW12} --- an improvement on the order of $10^4$ over MD in terms of computational speed for motions over comparable distances.
With {\sc Froda} we direct motion along a small number of low-frequency normal mode vectors \citep{Dia90,Hin98,SuhS04,PetP06,DobLS08}, the superposition of which well describes large-amplitude protein mobility \citep{KreAWE02,AleLEM05,JimFRW12}.  The resulting potential for {\sc Froda} to access the rarely sampled, high energy protein conformers  to which HDX is sensitive  \citep{SljW13} ensures that it is a natural choice of simulation method for improving our folding core predictions.

\section*{MATERIALS AND METHODS}

In addition to the FIR folding core,  we now introduce four other possible theoretical estimates for the folding core. These include information on the surface exposure and dynamics of the protein in addition to rigidity. The PDB structure 1CWA \citep{MikKPW93} was used for all simulations of protein rigidity and motion, before and after removal of the ligand CsA.
Initially we focus upon the CypA-CsA complex, before applying our analysis to the unbound CypA and comparing our results. 

\section*{A modified bond network for FIR$_B$}
\label{sec:HP_settings}

The presence and strength of HBs within the constraint networks were determined based upon donor-acceptor geometry using {\sc First}. Bonds weaker than an energy cutoff value $E_{\mathrm{cut}}$ are excluded from the network. During a rigidity dilution  $E_{\mathrm{cut}}$ is systematically lowered and HBs are consequently removed.\footnote{We emphasize that $E_{\mathrm{cut}}$ should be thought of as an arbitrary parameter and not as directly related to an energy scale for motion --- such as $k_\mathrm{B}T$ --- in the structure \cite{JimFRW12}.} 
In addition to the standard bond network defined in this way by {\sc First}, we will also consider a modified network based on the 
surface exposure of atoms with HBs and HPs. By demanding that both interacting atoms in a HB or HP constraint are buried within the protein, i.e.\ not exposed on the surface, we enhance the flexibility of the protein surface. The folding core generated in this way consists of residues which are both rigid and buried within the protein.
HPs are indirect, entropy-driven interactions  \citep{FolRD08} thought to contribute significantly to protein folding \citep{Dil90}.
HPs have not always been included in {\sc First} simulations, and their effect on modeling the flexibility and mobility of proteins is still not fully understood. In early papers using {\sc First}, HPs were not included in the bond network \citep{ThoHYK00,JacRKT01}, whereas more recently it is usual practice to include HPs and maintain them throughout rigidity dilution  \citep{ZavLTD04,RadG08}  or to increase their number as $E_{\mathrm{cut}}$ is lowered \citep{RatRG12,PflRKR13}.
In {\sc First}, HPs are modelled as flexible constraints, restricting the separation distance between atoms involved in the interaction, but not the angle between them. In this way, the interacting atoms are permitted to slip relative to each other \citep{HesRTK02,GohKC04}. 
HPs between carbon or sulfur atoms are typically included if the distance between these atoms is less than the sum of their van der Waals radii, $r_v$, plus a distance cutoff, $D_{\mathrm{HP}}$. For carbon and sulfur, $r_v = 1.7$~$\mathrm{\AA}$ and $1.8$~$\mathrm{\AA}$ respectively, and $D_{\mathrm{HP}}$ is typically set to 0.25~$\mathrm{\AA}$ \citep{HesRTK02,GohKC04,RadG08}. These distances will then allow to define the burial distance of a given atom (see below).

As a first improvement to FIR, we therefore modified the bond network in {\sc First}, requiring that each HP and HB is formed between two interacting atoms, both of which are buried within the protein. We implemented this bond network and performed a rigidity dilution on the structure 1CWA. As in Ref.\ \cite{HeaRBF14}, the folding core is determined from the dilution plot. In this case, the folding core consists of residues which are rigid \emph{and} buried, and we refer to this measure as FIR$_B$. 

\section*{Burial distances}
\label{sec:bur_method}

Let us now define in detail how we distinguish buried and exposed residues.
For each atom in the protein, a sphere of radius $r_v + r_w$ is drawn with the atom in the centre, where $r_v$ is the van der Waals radius of the atom and $r_w = 1.4$~\AA is that of a water molecule; for nitrogen, $r_{v}(N) = 1.5$~\AA\. 
The points forming the sphere are then individually checked for contact with the neighboring atoms. If any point on the sphere is not in contact with a neighbor then that is a potential solvent position and the atom is labeled as being exposed. If an atom is not exposed, it is buried and its burial distance, $R$, is the shortest distance to an exposed atom. An exposed atom has $R =0$~\AA.
For a given conformer, we calculated $R$ for each nitrogen atom in the amide backbone and used this  as the value of $R$ for the residue to which it belongs.

\section*{Coarse-grained mobility simulation for FRO}

\label{sec:Fr_method}
\label{sec:Pseudo_method}

The coarse-grained elastic network model implemented using the {\sc ElNemo} software \citep{SuhS04} allows for a rapid and accurate calculation of the normal modes of motion \citep{Tir96,Hin98}.  In Ref.\ \citep{WelJR09} we reported results for directed motion along these normal modes to generate new conformations satisfying the constraints of the {\sc First} bond network. 
Following this work, we bias motion along both (parallel and antiparallel) directions of the ten lowest frequency non-trivial normal mode vectors, $m_7$ -- $m_{16}$ with directed and random step sizes both set to 0.01~\AA. These low frequency modes well capture the large scale motion of a protein \citep{JimFRW12}. A total of $2000$ conformers were generated for each mode and direction. Details of the algorithm can be found elsewhere \citep{WelMHT05}. In each of our {\sc Froda} simulations, we use  $E_{\mathrm{cut}} = -2.0$ kcal/mol to determine the HB network. These simulations have been repeated at five different $E_{\mathrm{cut}}$ values ($-0.5, -1.0, -1.5, -2.0$ and $-3.0$~kcal/mol), but the nature of the results is not particularly sensitive to this (data not shown, see \citep{Hea13}).

We carried out simulations of motion using {\sc Froda} and calculated burial distances for all residues in the final (2000$^{th}$) conformer  from each of the {\sc Froda} simulations biased along a low frequency normal mode vector. The residues were classed as being buried ($R > r_v(N)\equiv 1.5$~\AA) or exposed ($R \leq 1.5$~\AA) in each of the conformers. Those which remained buried in more than half of the final conformations formed the {\sc Froda} folding core, which we refer to as FRO.

\section*{Secondary structure estimate SeS and the FIR+FRO combination F+F}

As well as FIR, FIR$_B$ and FRO introduced above, we consider two further theoretical folding cores, SeS and F+F. These are formed of the secondary structural units of helices and sheets (SeS) and the intersection of results between FIR and FRO (F+F). We choose to include SeS because of its simplicity and lack of need for any sophisticated computational analysis. The estimate F+F was included in order to test whether there is a distinct improvement upon combining data from static (rigidity, FIR) and dynamic (mobility, FRO) measures. 
In addition, we checked the intersection between FIR$_B$ and FRO. This is missing just one residue Val12 from F+F, but is otherwise identical. Since Val12 is part of the HDX folding core, the  FIR$_B$+FIR folding core makes a slightly weaker prediction than F+F in the present case.

\section*{Quantitative measures for comparing folding cores}

Let  $N_\mathrm{Ex}$ and $N_\mathrm{Th}$ be the number of residues contained in an experimentally and theoretically determined folding core, respectively. Clearly, $N_\mathrm{Ex}=N_\mathrm{Th}$ is a necessary condition for  agreement of experimentally and theoretically estimated folding cores. However, not only the number of residues, but of course the agreement of the specific set of residues in both experimental and theoretical folding cores is most important. In order to capture this, let us define ${\cal T}$ as the number of residues correctly identified by a theoretical prediction of the experimental folding core, i.e.\ ${\cal T}$ is the number of true positives. For a perfect agreement, we have ${\cal T}=N_\mathrm{Th}$ while the expected ${\cal T}$ attained randomly is $N_\mathrm{Th}N_\mathrm{Ex}/N$. Here $N$ denotes the total number of amino acids in the protein \citep{RadB04}.
We can define the \emph{specificity}, $\alpha$, and  \emph{sensitivity}, $\gamma$, of a theoretical folding core prediction as
\begin{equation}
\alpha = \frac{{\cal T}}{N_\mathrm{Th}}
\label{eqn:alpha}
\end{equation}
and
\begin{equation}
\gamma = \frac{{\cal T}}{N_\mathrm{Ex}}.
\label{eqn:gamma}
\end{equation}
The specificity $\alpha$ measures the fraction of residues identified by the theoretical method which are also part of the experimental folding core. The sensitivity $\gamma$ shows the proportion of residues in the experimental core which have been correctly predicted by the theoretical method. A perfect correspondence between theory and experiment, ${\cal T}=N_\mathrm{Th}=N_\mathrm{Ex}$, would yield $\alpha = \gamma = 1$ while for a completely wrong identification, ${\cal T}=0$, we have $\alpha = \gamma = 0$. 

Another, previously defined \citep{RadB04}, quantitative measure is the so-called folding core identification \emph{enhancement factor}, $\epsilon$.
This is the ratio of ${\cal T}$ and the number of residues expected to be identified by a random selection,
\begin{equation}
\epsilon = \frac{{\cal T} N}{{N_\mathrm{Th}  N_\mathrm{Ex}}}.
\label{eqn:epsilon}
\end{equation}
\label{eqn:s(AB)}
A theoretical method with random probability of success has $\epsilon = 1$, and when $\epsilon > 1$, the match is better than random. We shall also compare our $\alpha$ and $\gamma$ measures to $\epsilon$.

\section*{RESULTS AND DISCUSSION}

\section*{Effects of the modified bond network and the inclusion of mobility}

\begin{figure}[tbp]
\begin{center}
\includegraphics[width=\columnwidth]{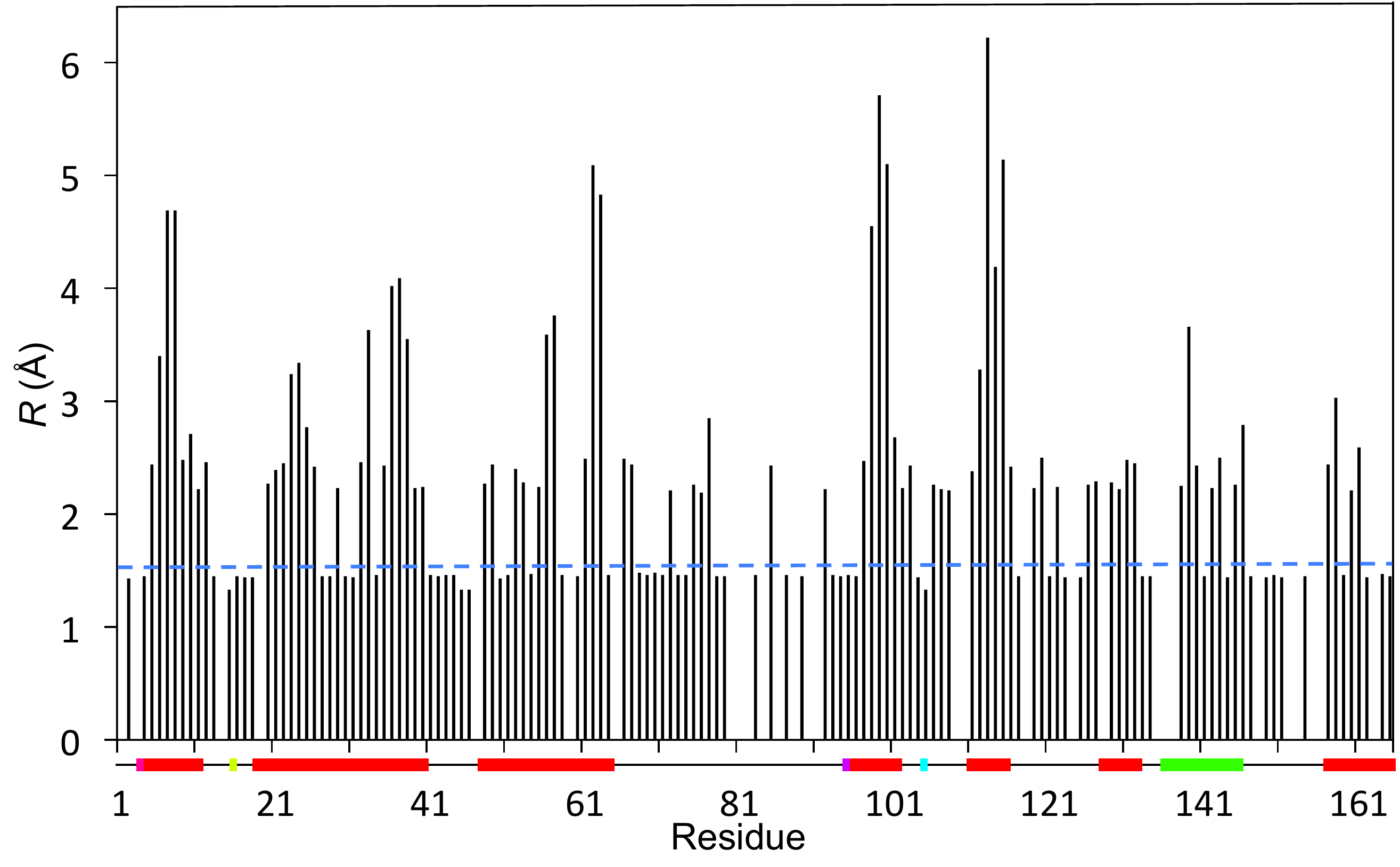}
\caption[]{The burial distance $R$ for each residue of CypA in the static structure of the CypA-CsA complex. The blue dashed line at $R = 1.5$~\AA\ is used to separate exposed residues (below the line) from buried residues (above the line). Below the horizontal axis, we schematically show the extent of the FIR$_B$ folding core along the residues via coloured regions. Those residues which are part of FIR$_B$ are coloured red; other rigid residues are coloured otherwise. The thin horizontal black line denotes residues that are flexible.}
\label{fig:stat_bur}
\end{center}
\end{figure}

We measured the surface exposure of each amide nitrogen in the structure 1CWA, cp.\ Figure \ref{fig:stat_bur}. For the small protein CypA, we find that the majority of the residues are somewhat exposed to the protein surface, with $R<r_v(N)=1.5$~\AA. Since we measure the burial of the amide nitrogen and the length of the N-H bond is $1.5$~\AA, if the H atom is exposed to the surface, then the amide nitrogen has $R \leq 1.5$~\AA.
In Figure~\ref{fig:stat_bur} we also show a superposition of FIR$_B$ along the protein backbone as taken from the rigidity dilution plot. We note that most of the residues that are highly buried within the protein structure ($R\gtrsim 2.0$~\AA) correspond to regions of the protein that are part of the folding core. This shows that the majority of those residues that are buried in the CypA-CsA complex are in fact rigid \emph{and} buried. 

\begin{figure}[tbp]
\begin{center}
\includegraphics[width=\columnwidth]{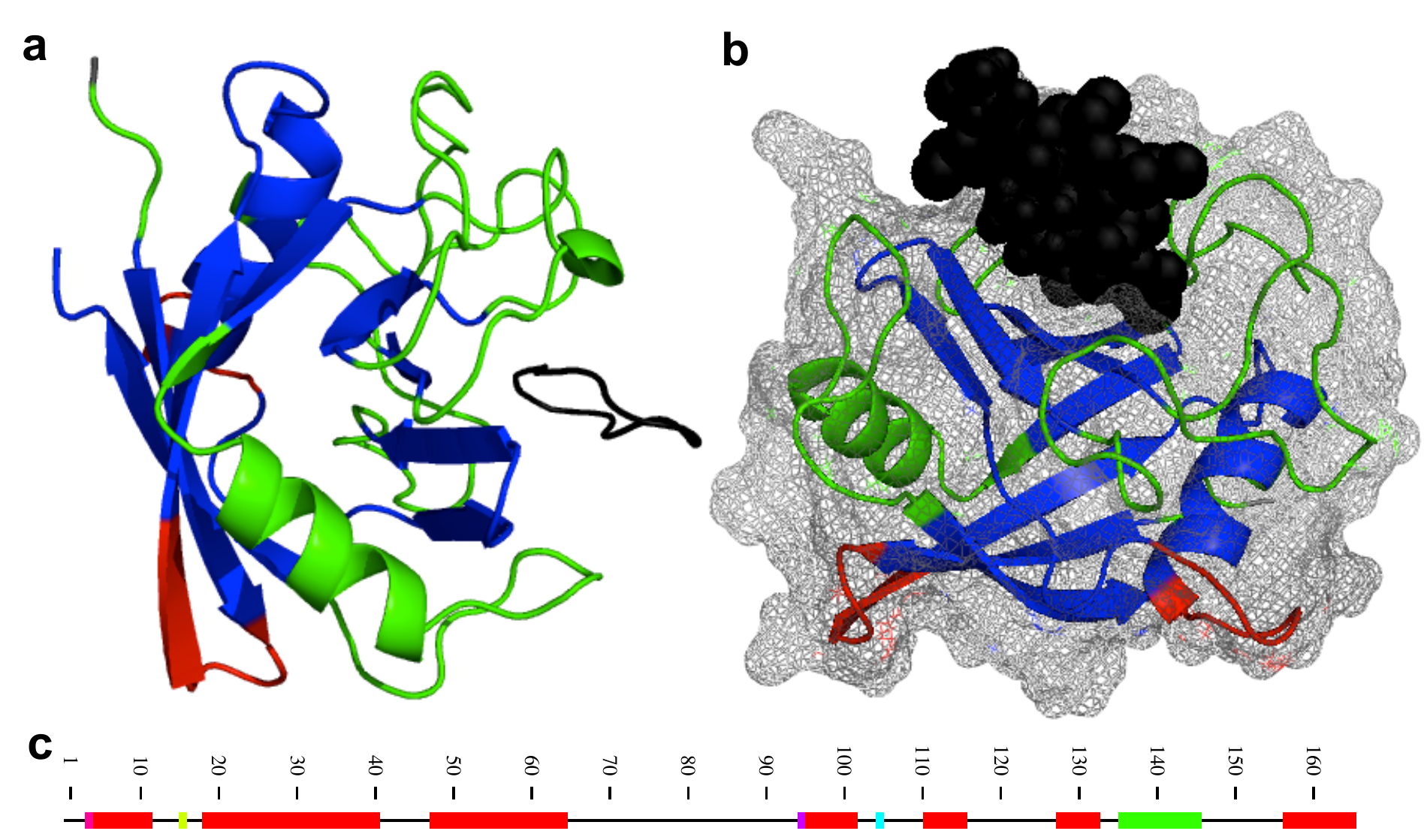}
\caption[]{(a) {\sc PyMOL} cartoon of 1CWA with FIR$_B$ in blue and the ligand, CsA, in black. Those residues which are part of FIR but not FIR$_B$ are colored red, with the other residues of CypA in green (the colour scheme is similar to the one used in Fig.\ \ref{fig:stat_bur}). (b) The surface of the protein is illustrated as a mesh, with CsA represented as spheres. The image has been rotated from its orientation in (a) by $180^{\circ}$ about the $45^\circ$ diagonal. (c) A linear representation  by residue number of the rigid cluster decomposition at $E_{\mathrm{cut}} = -1.061$~kcal/mol shown as thick colored blocks (rigid) or thin black lines (flexible). The mutually rigid red blocks show the amino acids  which constitute FIR$_B$.} 
\label{fig:HB0_FC}
\end{center}
\end{figure}

Let us now see how our improved FIR$_B$ folding core estimate is different from the FIR folding core of Ref.\ \cite{HeaRBF14}. Figures \ref{fig:HB0_FC} (a) and (b) show a cartoon representation of the residues identified as belonging to the FIR$_B$ folding core superimposed on the 3D structure of the protein. We also indicate in Fig.\ \ref{fig:HB0_FC} (c) the FIR$_B$ folding core along the protein chain as in Figure~\ref{fig:stat_bur}. This is similar to the cartoon representation of the comparison between the HDX folding core and the FIR estimate given in Ref.\ \cite{HeaRBF14}.
As expected, FIR$_B$ is smaller than FIR, since there are fewer constraints present throughout the rigidity dilution. The amino acids $12$--$18$ and $41$--$47$ are part of FIR but not part of FIR$_B$. These residues are close to the protein surface on the opposite side of the protein to the ligand binding site as can be seen in Figure \ref{fig:HB0_FC}(b).

\begin{figure*}[bht]
\begin{center}
\includegraphics[width=\textwidth]{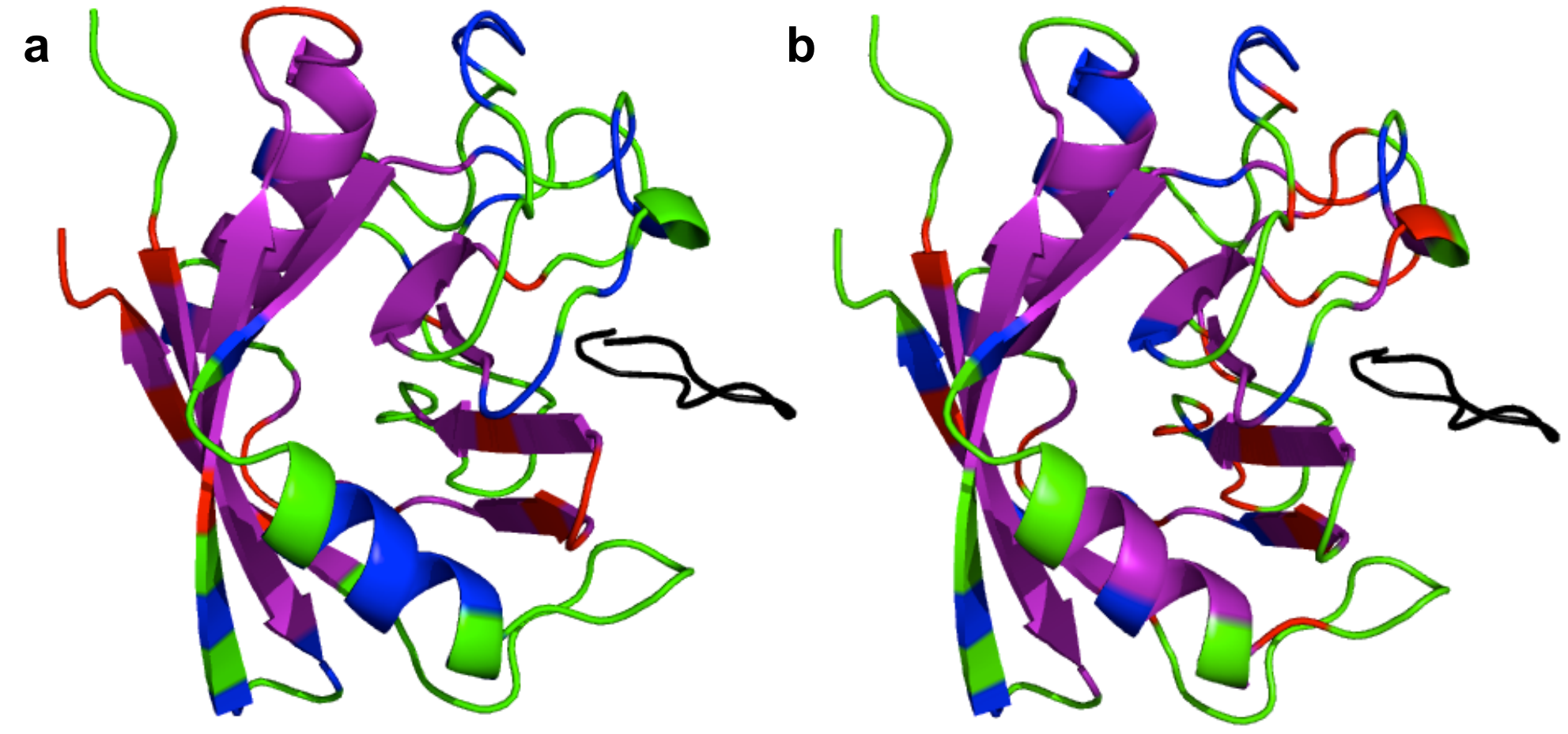}
\caption[]{Theoretical folding cores (a) FIR$_B$ and (b) FRO for the CypA-CsA complex compared with the HDX folding core. In both cases, residues are colored purple (part of both folding cores),
      blue (HDX folding core only), red (theoretical folding core
      only) or green (not part of either folding core). CsA is
      colored black.}
\label{fig:froda_FC}
\end{center}
\end{figure*}

Figure \ref{fig:froda_FC}(a) shows a qualitative comparison between FIR$_B$ and the HDX folding core, depicted on the 3D structure of the CypA-CsA complex.
We find that for most of the residues in the protein, the theoretical and experimental folding cores are in agreement. However, residues 138 -- 143, forming the $\alpha$-helix at the forefront of the image, are part of the HDX folding core, but are \emph{not} part of FIR$_B$. With reference to the representation of FIR$_B$ along the protein chain, as shown in Figure \ref{fig:HB0_FC}(c), we see that these residues are indeed rigid at $E_\mathrm{cut}$ corresponding to FIR$_B$, but they are not part of the largest rigid cluster. In this case, loosening the definition of the {\sc First} folding core to include residues that are rigid independently of the largest rigid cluster would capture this slowly exchanging helix as part of the rigid folding core. 

In Figure \ref{fig:froda_FC}(b) we show the {\sc Froda} folding core compared qualitatively with the HDX folding core. 
Notably, residues $138$ -- $143$ in the helix are now part of the FRO folding core (but not of FIR$_B$). We emphasise that the results of this mobility-based analysis depend indirectly on the results of the rigidity analysis; {\sc First} identifies a constraint network and {\sc Froda} then explores the motion that is possible within those constraints. Therefore it is possible for a residue to be (i) rigid but exposed --- thus lying in FIR or FIR$_B$ but not in FRO --- or (ii) non-rigid but well protected by burial within the protein --- thus lying in FRO but not in FIR or FIR$_B$.
This suggests that in order to predict the results of the HDX folding core, which depend upon surface exposure as well as dynamics (inherited from flexibility) a combination of the approaches employed by {\sc First} and by {\sc Froda} may be necessary. 

\begin{figure*}[tbp]
\begin{center}
\includegraphics[width=\textwidth]{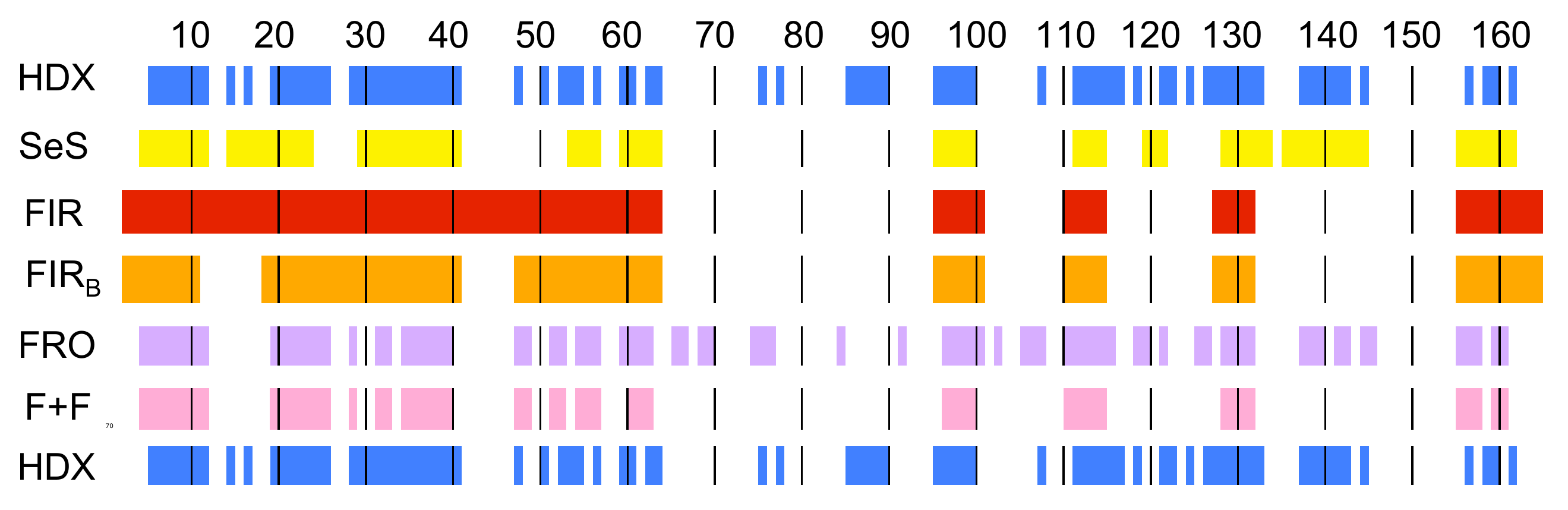}
\caption[Theoretical and experimental folding cores]{Comparison of the experimentally determined HDX folding core for CypA-CsA (blue and shown above and below the other folding cores) with the five folding cores computed using the theoretical approaches outlined in the text and labeled accordingly (from top to bottom: SeS (yellow), FIR (red), FIR$_\mathrm{B}$ (orange), FRO (purple), and F+F (pink)). 
In each case the folding core residues in the primary structure of CypA are represented as colored blocks while non-folding core residues are kept white. The residue
      numbers along the protein backbone are indicated horizontally, with thin black vertical lines
      added every ten residues for clarity.}
\label{fig:FCs}
\end{center}
\end{figure*}

\begin{figure*}[tbp]
\begin{center}
\includegraphics[width=\textwidth]{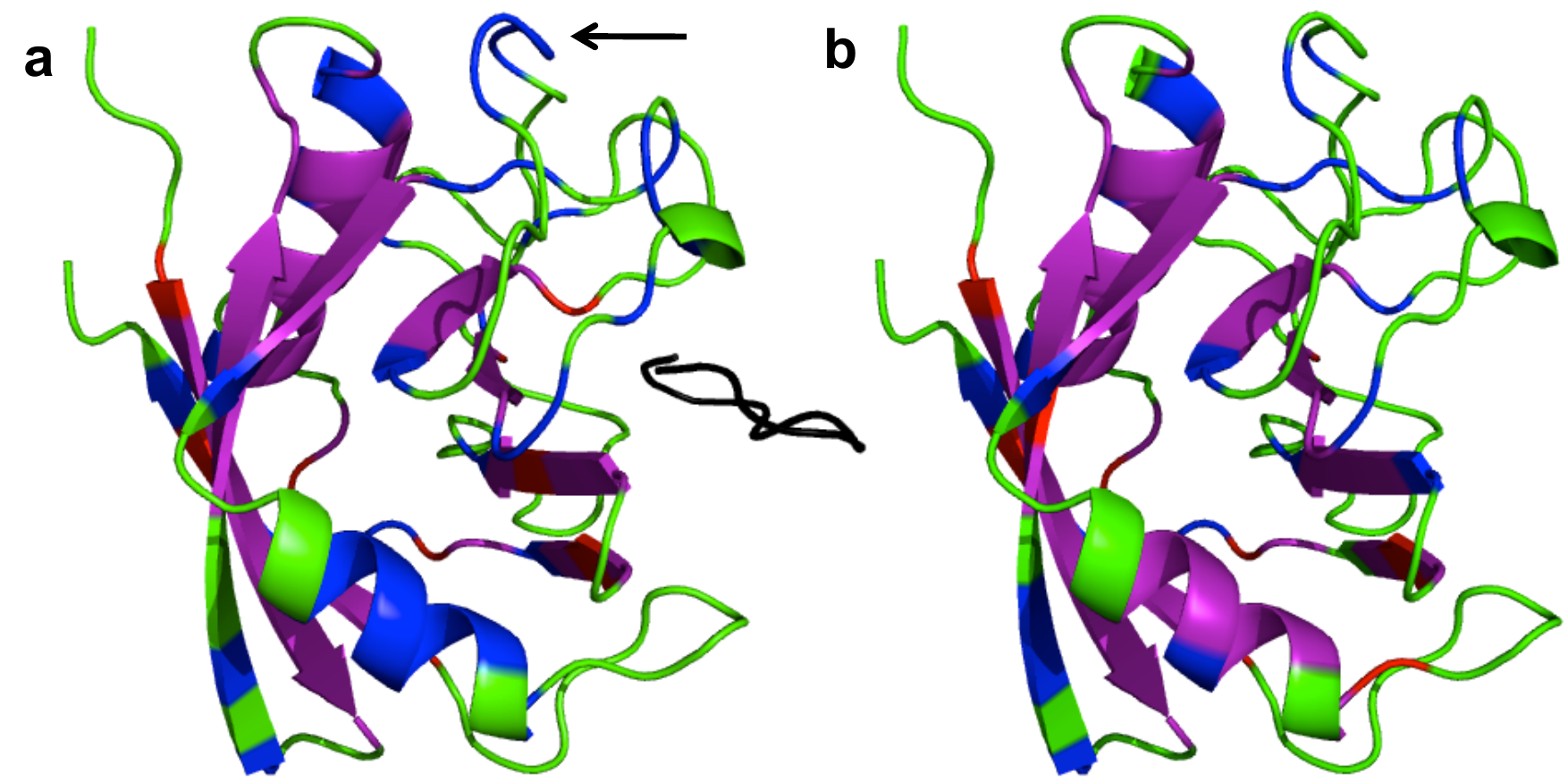}
\caption{Cartoon representation of a comparison of HDX folding cores with F+F for (a) the CypA-CsA complex and (b) unbound CypA. In both cases, residues are colored purple (part of both folding cores), blue (HDX folding core only), red (F+F folding core
      only) or green (not part of either folding core). CsA is
      colored black in (a). The arrow in (a) indicates residues $86$--$90$.}
\label{fig:F+F_FC}
\end{center}
\end{figure*}

In Figure \ref{fig:FCs}, each of the five theoretical folding cores are compared with HDX folding cores for the CypA-CsA complex. In each case, folding core residues along the protein backbone are shown as bold blocks of colour.
FIR, FIR$_B$ and FRO have been discussed above, so we here briefly discuss the remaining folding cores definition SeS and F+F. 
We find that SeS consists almost entirely of slowly exchanging amino acids, although there are also many slowly exchanging residues which are not part of a secondary structure unit. 
The residues in FRO are also for the most part slowly exchanging.
We note that some parts of the HDX folding core, for example residues $86$--$90$, are poorly predicted by all of the methods. This region corresponds to a flexible, surface-exposed, unstructured region in the crystal structure, indicated by an arrow in Figure \ref{fig:F+F_FC}(a).
Our methods, based around predicting rigid, buried and immobile regions, fall short of identifying this area as slowly exchanging. 
Figure \ref{fig:F+F_FC}(b) presents a graphical comparison between the HDX folding core and F+F on the structure of CypA.  We observe that in the absence of the ligand, residues  $86$--$90$ are mostly absent from both theoretical and experimental folding cores. The relatively small number of false positives (red) to false negatives (blue) demonstrate the high ratio of specificity to sensitivity for this method.

\section*{Quantitative analysis of theoretical folding cores}

Table \ref{tab:sAB} shows the number of residues in each theoretical folding core along with the number of true positives ${\cal T}$.  The comparison pictured in Figure \ref{fig:FCs} is made quantitative with the specificity $\alpha$, the sensitivity $\gamma$, and the enhancement factor $\epsilon$, as defined in Equations (\ref{eqn:alpha}), (\ref{eqn:gamma}) and (\ref{eqn:epsilon}), respectively. 
\begin{table*}[bt]
\caption[Quantitative measures of agreement between theoretical methods and HDX folding core for CypA-CsA and unbound CypA]{Quantitative measures of agreement between theoretical folding cores and the experimental HDX folding core for (top) CypA-CsA  and (bottom) unbound CypA with $N=165$. The bold numbers indicate the largest selectivity $\alpha$, sensitivity $\gamma$ and enhancement $\epsilon$ for CypA-CsA and unbound CypA. The error estimates for $\alpha$ and $\gamma$ reflect an assumed $\pm 5\%$ variation in $N_\mathrm{Ex}$, $N_\mathrm{Th}$ and ${\cal T}$. The two estimates for the enhancement $\epsilon$ show the variation when using the full set of residues with $N=165$ and when excluding the $6$ proline residues of CypA such that $N=159$. 
}
\begin{center}
\begin{tabular}{|l|c|c|c|c|c|c|c|c|}
\hline
& Folding core & $N_\mathrm{Ex}$ & $N_\mathrm{Th}$ & ${\cal T}$  & $\alpha$ & $\gamma$ & $\epsilon_{165}$  & $\epsilon_{159}$ \\
\hline
\hline
& SeS      & 80 &	75	&	59	 &	$0.79\pm 0.04$ 	&	{$\mathbf{0.74}\pm 0.04$} &	$1.62$	& $1.56$	\\
& FIR	& 80 &	86	&	57	 &	$0.66\pm 0.04$ 	&	$0.71\pm 0.04$ &	$1.37$	& $1.32$	\\
CypA-CsA
& FIR$_B$ & 80 &	72 	&	53	 &	$0.74\pm 0.04$ 	&	$0.66\pm 0.04$  &	$1.52$	& $1.46$ \\
& FRO	& 80 &	80	&	56	 &	$0.70\pm 0.04$ 	&	$0.70\pm 0.04$ &	$1.44$	& $1.39$	\\
& F+F	& 80 &	54	&	44	 &	$\mathbf{0.81}\pm 0.05$ & $0.55\pm 0.03$ &	$\mathbf{1.68}$	 & $\mathbf{1.62}$	\\
\hline
& SeS	& 73 & 75	& 58  & $0.77\pm 0.04$ & $0.79\pm 0.04$ & $1.75$ & $1.68$  \\
& FIR 	& 73 & 112	& 66  & $0.59\pm 0.03$ & $\mathbf{0.90}\pm 0.05$ & $1.33$ & $1.28$ \\
unbound CypA
& FIR$_B$ & 73 & 72 	& 52 & $0.72\pm 0.04$ & $0.71\pm 0.04$ & $1.63$ & $1.57$  \\
& FRO 	& 73 & 76 	& 54  & $0.71\pm 0.04$& $0.74\pm 0.04$  & $1.61$ & $1.54$\\
& F+F 	& 73 & 59 	& 51 & $\mathbf{0.86}\pm 0.05$ & $0.70\pm 0.04$ & $\mathbf{1.95}$ & $\mathbf{1.88}$  \\
\hline
\end{tabular}
\end{center}
\label{tab:sAB}
\end{table*}%
%
Each of the five theoretical folding cores are significant improvements upon random selections, as demonstrated by the values of $\epsilon$, the lowest of which is $1.33$. The four theoretical methods introduced in this paper all outscore FIR with respect to this enhancement measure, with F+F scoring highest for both the CypA-CsA complex and the unbound CypA, with SeS second. SeS is highly selective and sensitive in both cases, with specificity $\alpha \geq 0.77$ and sensitivity $\gamma \geq 0.74$. Unfortunately, the simple SeS method of folding core selection is clearly inappropriate for predicting changes upon ligand binding, since by definition it does not change.
As discussed in Ref.\ \cite{HeaRBF14}, FIR is sensitive to the bond network and its increase in size upon ligand removal is in disagreement with the experimental results, which show a larger folding core in the presence of CsA. The most accurate method for CypA-CsA and unbound CypA is F+F, with $\alpha = 0.81$ and $0.86$.  This high specificity comes at the cost of lower sensitivity, meaning that it predicts with high accuracy but does not capture all of the HDX folding core. A larger proportion of the HDX folding core needs to be accurately determined by this method in order to successfully capture the impact of ligand binding on HDX.
We have also estimated the variation in $\alpha$, $\gamma$ and $\epsilon$ in Table \ref{tab:sAB}, assuming a $\pm 5\%$ error in $N_\mathrm{Ex}$, $N_\mathrm{Th}$ and ${\cal T}$. This leads to some variation in the results for $\alpha$ and $\gamma$ and allows us to estimate the accuracy of the numbers quoted here. In addition, due to their lack of an amide proton, the six proline residues in CypA do not appear in the $^1$H-$^{15}$N HSQC spectrum and so cannot be part of the HDX folding core. For this reason, we also take $N=159$ when calculating $\epsilon$. We find that in most cases the largest values of $\epsilon$, $\alpha$ and $\gamma$ remain significantly larger. This shows that our findings are indeed robust.

\section*{Effect of ligand binding on folding cores}

\begin{figure}[htbp]
\begin{center}
\includegraphics[width=\columnwidth]{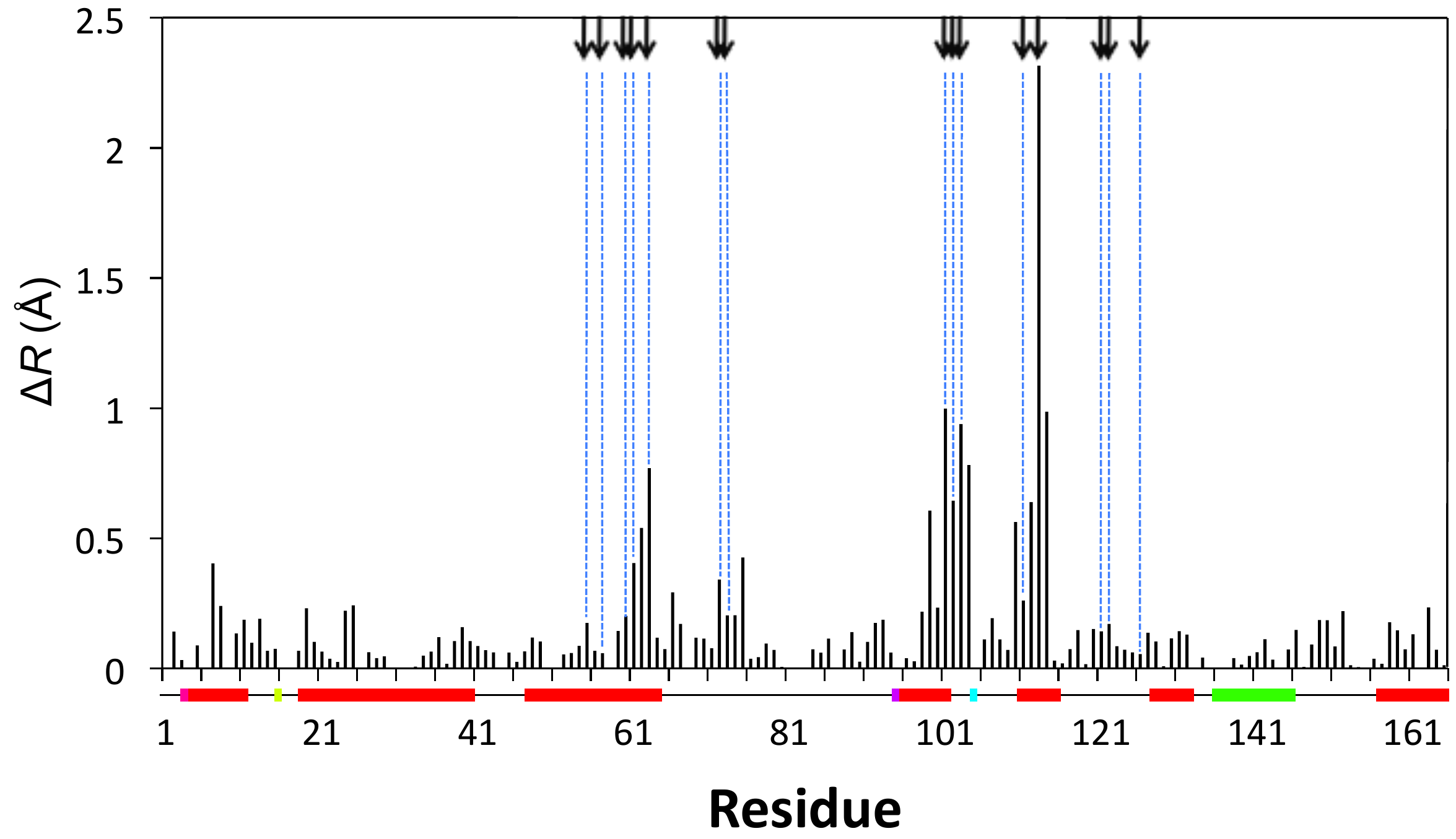}
\caption[Ligand binding affects $R$ for binding site residues]{The difference in burial distance $\Delta R$ (black bars) between the CypA-CsA complex and the unbound protein plotted for each amino acid. The arrows and vertical dotted lines indicate the binding residues. As in Figures \ref{fig:HB0_FC} and \ref{fig:stat_bur}, FIR$_B$ is also shown.}
\label{fig:burial_abs}
\end{center}
\end{figure}

As demonstrated in Ref.\ \cite{HeaRBF14}, for CypA there are only small changes to the HDX folding core upon ligand binding, and these are not captured using rigidity analysis alone. Indeed, only seven residues are part of the HDX folding core for the complex but not for the unbound protein.
Upon ligand removal, FIR$_B$ changes by a single residue, which is not one of these seven. As discussed in Ref.\ \cite{HeaRBF14}, the FIR changes unexpectedly upon ligand binding. These two folding cores solely derived using {\sc First} do not reflect the effect observed experimentally. We see a larger effect of ligand binding on FRO.
Figure \ref{fig:burial_abs} shows how ligand binding affects the burial distance $R$. We calculated the average $R$ for each residue from the final conformers of the {\sc Froda} simulations, and plotted the absolute difference, $\Delta R$, between these values.
Of the eleven residues for which $\Delta R > 0.5$ $\mathrm{\AA}$, five are binding site residues and none are further than two residues along the backbone from a binding site residue.
We also calculated $\Delta R$ for each residue  from {\sc Froda} simulations carried  out at $E_{\mathrm{cut}}$ = $-0.5$, $-1.0$, $-1.5$ and $-3.0$ kcal/mol.
In each case, the residues which are part of, or close to, the binding site have the largest $\Delta R$ (data not shown).
 The HDX folding core changes in sections of the protein more distant from the binding site, and so tracking the protein surface in {\sc Froda} does not capture this effect.

\section*{CONCLUSION}
We have introduced and discussed rapid computational methods for adding information on surface exposure and protein dynamics into a rigidity-based folding core definition. These improve upon the {\sc First} folding core for predicting the HDX folding cores of both CypA-CsA and unliganded CypA. In addition, we show that they are less sensitive to the erroneous increase in folding core size upon ligand removal observed using {\sc First} with the standard bond network. On balance, F+F, which combines rigidity and motion, appears as the best choice for a computational determination of the folding core from a protein's structural information alone.
Nevertheless, no method achieves a perfect score of $\alpha=\gamma=1$, neither for the uncomplexed CypA nor for the CypA-CsA complex. This insensitivity to ligand binding is disappointing but unsurprising due to the scale of the challenge --- for CypA-CsA and in general.
 
We have sought to obtain a practical balance between the computational demands of our predictions and their utility. The requirement for large-scale motions make more computationally sophisticated methods such as MD inappropriate. Methods using simpler input data than ours, making sequence-based predictions of surface exposure and then in turn predictions of HDX folding cores, are currently unsatisfactory. We feel that although balance between sophistication of input data and method and the accuracy of the prediction is yet to be struck, it may be achieved through coarse-grained approaches. The {\sc froda} alternative {\sc frodan} has more flexibility of hydrogen bond and hydrophobic contact geometry built in to its bond networks. A comparison between the predictions made using {\sc frodan} instead of {\sc froda} would be a valuable extension to this work.  

Our overall strategy is based on using the results from HDX experiments as the benchmark against which theoretical  results are compared. The measures $\alpha$, $\gamma$ and $\epsilon$  score most highly when the match of a theoretical folding core with the HDX result is perfect. However, the HDX data may contain errors themselves and this additional source of variation did not play a role in our judgements of the success of the computational approaches. 
Furthermore, for the CypA-CsA complex, the protein is large relative to the ligand and the ligands effect on HDX is rather small.
We expect that an application of our methods to a larger protein that exhibits a more significant conformational change upon ligand binding will also give a clearer change in the values of $\alpha$, $\gamma$ and $\epsilon$.

%

\section*{ACKNOWLEDGMENTS}

\ack{We thank C.\ Blindauer for help with the experimental NMR aspects of this work. The CypA plasmid was kindly provided by G.\ Fischer from the Max Planck Research Unit for Enzymology of Protein Folding in Halle, Germany. We gratefully acknowledge funding from the EPSRC Life Sciences Interface programme (MOAC DTC EP/F500378/1). JWH thanks the Institute of Advanced Study in Warwick for an Early Career Fellowship; SAW is grateful to the Leverhulme Trust for an Early Career Fellowship.}\vspace*{6pt}



\bibliographystyle{biophysj}\bibliography{bibliograph}

\end{document}